\documentclass[prl,twocolumn,superscriptaddress,]{revtex4}
\usepackage{dcolumn}
\usepackage{bm}
\usepackage{amssymb}
\usepackage{graphicx,epsfig}
\begin{document}

\title{Comment on the paper "Semiclassical Dirac Theory of Tunnel
Ionization"\\ {\normalsize by N.Milosevic, V.P.Krainov, and
T.Brabec}}

\bigskip
\author{ B. M. Karnakov\thanks{corresponding author.\\
 E-mail address:karnak@theor.mephi.ru (B.M.Karnakov}}
\author{V. D. Mur}
\affiliation{Department of Theoretical Nuclear Physics,\\ Moscow
Engineering Physics Institute, Moscow 115409, Russia}
\author{V. S. Popov}
\affiliation{Institute of Theoretical and Experimental Physics,\\
 Moscow 117218, Russia}

\maketitle

In the paper \cite{1} the equation (8) for the ionization rate
$w_r$ of the ground state $1s_{1/2}$ of the hydrogen-like atom by
a constant crossed electromagnetic field was given,
$$ w_r=\frac{(eF)^{1-2\epsilon}}{2\sqrt{3}\xi
\Gamma(2\epsilon+1)}\sqrt{\frac{3-\xi^2}{3+\xi^2}}\Biggl
(\frac{4\xi^3(3-\xi^2)^2}{\sqrt{3}(1+\xi^2)}\Biggr
)^{2\epsilon}\times$$
$$\times \Biggl (6\mu~\arcsin
\frac{\xi}{\sqrt{3}}-\frac{2\sqrt{3}\xi^3}{eF(1+\xi^2)} \Biggr
).\eqno(*)
$$
This equation is considered by the authors of Ref. \cite{1} as the
main result of the paper, which gives "for the first time a
quantitative description of tunnel ionization of atomic ions" or
"the first quantitative determination of tunnelling in atomic ions
in the relativistic regime" (see the text before Eq.(9)  and
Abstract in \cite{1}).

In connection with this claim we consider it necessary to note the
following. The main results of Ref.\cite{1} are not original but
have been taken from previous works \cite{2,3,4,5}, published much
earlier than paper \cite{1}. The results \cite{2,3,4,5} ate well
known to the authors of Ref.\cite{1}
(they have referred to these results). Indeed Eq.$(*)$ can be
easily rewritten in the following form:
\begin{equation} \label{wr}
w_r=\frac{m_e c^2}{\hbar}C_\lambda^2~ P~ \tilde{Q}~ Exp,
\end{equation}
where the exponential term $Exp$ and the preexponential term $P$
are identically equal to the expressions obtained
 in Refs.\cite{2,3,4,5}
\begin{equation}
Exp=\exp\left(-\frac{2\sqrt{3}~\xi^3}{1+\xi^2} \frac{{\cal
E}_{S}}{{\cal E}}\right),~~~P=\frac{1}{\xi}
\sqrt{\frac{1-\xi^2/3}{3+\xi^2}}\frac{{\cal E}}{{\cal{E}}_S},
\end{equation}
see, in particular, Eqs.(17),(35) and (50) in Ref.\cite{4}.
Comparing these formulae  with Eq.$(*)$  one should take into
account the notations: $$e F \equiv {\cal{E}}/F_{cr},~ F_{cr}
\equiv {\cal{E}}_S = 1/e,~~ \mu = Z\alpha = Z/137$$ (in
relativistic units $\hbar = m = c = 1$). Here ${\cal{E}}$ is the
electric field strength, ${\cal E}_S=m_e^2 c^3/e\hbar$ is the
Schwinger field in QED \cite{6}, and $\xi$ is a convenient
auxiliary variable introduced in Ref.[3].  This variable is
natural  for the problem of subbarrier electron motion in the
framework of the "imaginary time" method [3,4]:
\begin{equation}
\xi=\left[1-\frac{1}{2}\varepsilon\left(
\sqrt{\varepsilon^2+8}-\varepsilon\right)\right]^{1/2},
\end{equation}
where $\varepsilon = E_0/m_ec^2$ is the reduced energy of the
initial atomic state.

The factor $\tilde{Q}$ in Eq.(1) takes into account the Coulomb
interaction between the outgoing electron and the atomic core and
was calculated in the framework of the quasiclassical perturbation
theory [7]. It exactly coincides with our factor $Q$,
\begin{equation}
Q=\Biggl [\frac{2\xi^3(3-\xi^2)^2}{\sqrt{3}(1+\xi^2)}\cdot
\frac{F_{cr}}{\cal{E}}\Biggr ]^{2\eta}\cdot \exp \Biggl(6Z\alpha~
\arcsin\frac{\xi}{\sqrt{3}}\Biggr ),
\end{equation}
$$ \eta = Z \alpha \varepsilon/ \sqrt{1 - \varepsilon^2}$$ (see
Eq.(35) in Ref.\cite{4}), in the particular case of the
$1s_{1/2}$ state of a hydrogen-like atom with the nuclear electric
charge  $Z$, when
\begin{equation}
\varepsilon\equiv  \eta = \sqrt{1-(Z\alpha)^2},~~~\tilde{Q}\equiv
Q.
\end{equation}

However, it is correct only in the case when there is only one
electron in the $K$-shell of atom, and all other electrons are
turned off. Our formula (4)  is much more general, because it is
applicable in the case of atomic ions with an arbitrary degree of
ionization, if the parameters $\varepsilon$ and $C_\lambda^2$ are
taken from the independent calculations for the case of atoms
without external fields (for example, in the framework of
Hartree-Fock-Dirac method), or directly from the experimental
data.

The factor $C_\lambda$ is the asymptotic coefficient of the atomic
wave function at large distances from the nucleus. In Ref.\cite{1}
this factor was assumed to be equal to the text-book
value corresponding to the $1s_{1/2}$ state, see, for example,
Eq.(14.39) in Ref.\cite{8}. In the general case $C_{\lambda}$ is
an independent parameter and should be determined numerically.

The majority of the formulae given in Ref.\cite{1}, including
"the main result", are literal reproductions of the corresponding
formulae from our papers \cite{3,4,5}. All the notations are the
same, including even the transition from energy $\varepsilon$ to
variable $\xi$, Eq.(3). The original contribution of the authors
of Ref.\cite{1} is entirely reduced to the trivial multiplication
of factors $Exp$, $Q$, and $P$, which were obtained in the
previous papers \cite{2,3,4,5}. However, Eq.$(*)$ which was
obtained as a result  of this operation, has nothing to do with
Dirac equation since according to it the tunnelling probability
does not include the spin factor which  appears in the process of
the subbarrier propagation of the electron.

Please note that the relativistic generalization of the Keldysh
ionization theory \cite{9}
was considered for the first time by Nikishov and Ritus \cite{2} for
spinless particles. In the case of constant
crossed fields the formulae  obtained by Nikishov and Ritus  coincide
with Eqs.(2), if they are expressed via the variable $\xi$.
The reference [11] in \cite{1} shows that our papers are known to
the authors of Ref.\cite{1}. However, our papers are referred
 as "an
analytical solution of  the Klein-Gordon equation for $\pi^-$
atoms in static electric and magnetic fields," though we did not
mention $\pi^-$ atoms at all.  Contrary to that, it is repeatedly
stressed in Refs. \cite{3,4,5} that we are aiming at the extension
of the imaginary time method to the case of relativistic
sub-barrier electron motion and its application in the theory of
deep-lying state ionization (including the $K$-shell) in heavy
atoms. In order to carry out experiments of $\pi^-$-atom
ionization the field strength should be increased by 5 orders of
magnitude (because ${\cal E}_S\sim m^2$). Correspondingly,
intensity $J$ of the laser beam will then reach the fantastic
value of $J\ge 10^{32}$ W/cm$^2$, which can hardly be achieved
because of the electron-positron pair production from vacuum by
the external electric field (the so-called Schwinger effect [6]).

Finally, we have to state that the paper \cite{1} is just a
compilation of our papers \cite{3,4,5}, written and published  few
years earlier. Thus it
 should not be published in  research journal,
such as  Physical Review Letters,  and in any case the authors
should point out that the formulae they have given had been
obtained earlier in Refs.\cite{2,3,4,5}, but not to pretend that they had
obtained them ``for the first time''.
All of the above can also be said about a more  detailed paper of the
same  authors \cite{10},  where the materials from \cite{1} are
supplemented with formulae of relativistic theory of ionization by
a static electric  field, completely taken from the earlier paper
\cite{4}. For example, compare Eq.(35) in \cite{10} with identical
Eqs.(6), (32) from Ref.\cite{4}.

The authors would like to gratefully acknowledge Professors S.V.Bulanov,
L.V.Keldysh,  N.B.Narozhny, L.B.Okun, Yu.A.Simonov and M.I.Vysotsky
for the discussion of this letter, valuable remarks and
countenance as well as  S.S.Bulanov and N.S.Libova for their help
in translating of the manuscript.

\end{document}